\documentclass[prc,aps,preprint,showpacs]{revtex4}
\usepackage{graphicx}

\begin{document}
\title{Examining possible neutron-halo nuclei heaver than $^{37}$Mg} 

\author{ Ikuko Hamamoto$^{1,2}$ }

\affiliation{
$^{1}$ {\it Riken Nishina Center, Wako, Saitama 351-0198, Japan } \\ 
$^{2}$ {\it Division of Mathematical Physics, Lund Institute of Technology 
at the University of Lund, Lund, Sweden} }   




\begin{abstract}
The even-Z odd-N neutron-halo nuclei, which are the possible lightest
neutron-halo nuclei heavier than 
$^{37}$Mg, are explored by studying the shell-structure unique in 
weakly-bound neutrons for spherical or deformed shape.  
It is pointed out that 
due to the narrowed N=50 spherical energy-gap and a few resulting  
close-lying neutron
one-particle levels, 1g$_{9/2}$, 3s$_{1/2}$, and 2d$_{5/2}$,   
for spherical shape, nuclei with some weakly-bound neutrons filling 
in those levels may be
deformed and have a good chance to show deformed s-wave halo. 
Promising candidates are $^{71}_{24}$Cr$_{47}$, $^{73}_{24}$Cr$_{49}$, 
$^{75}_{24}$Cr$_{51}$ and $^{77}_{26}$Fe$_{51}$ in the case that 
those nuclei lie inside the
neutron drip-line.   An interesting possibility of the deformed p-wave or s-wave
halo is suggested also for the nucleus $^{53}_{18}$Ar$_{35}$.
\end{abstract}


\maketitle

\newpage

\section{INTRODUCTION} 
Halo phenomena, on which the experimental information started from Ref. 
 \cite{IT85}, are so far observed mostly in lighter nuclei 
close to neutron-drip-line and are 
one of the most interesting phenomena in nuclei away from the stability line.  
In order to detect the occurrence of neutron halo, 
either the measurement of the larger 
mean-square-radius of the matter distribution or the detection of 
an appreciable amount of the probability of  
(one or several) weakly-bound neutrons far away from well-bound core nuclei 
is most efficient.   The
typical experiment done so far to measure the former quantity  
is the measurement of reaction cross sections ($\sigma_{R}$) \cite{MT14},
while the typical experiment to efficiently detect the latter one 
is to observe Coulomb break-up reactions \cite{TN09}.    

To my knowledge, the nucleus $^{37}_{12}$Mg$_{25}$ is so far the heaviest
nucleus which showed (deformed p-wave) halo properties \cite{NK14}.   
One may wonder which
nucleus heavier than $^{37}$Mg may be expected to show (spherical or deformed) 
halo.  In heavier nuclei it will be getting more and more difficult to clearly 
identify halo phenomena from the values of 
root-mean-square radius, because the
ratio of the number of halo particles to that of nucleons in 
well-bound core nucleus gets
smaller.   Furthermore, some amount of increase of the matter radius 
may occur also by the
change from spherical to deformed shape.  And, 
the rate of the increase of the matter radius due to possible
deformation may not generally decrease in
heavier nuclei.   On the other hand, the method such as Coulomb break-up
reactions can be efficient in discovering possible halo also in heavier nuclei.    
  
It is noted that the analysis of the data on Coulomb and nuclear 
break-up reactions of
$^{37}$Mg so far 
seems to show that the
experimental data can be well understood in terms of particle
picture without introducing the effects of pair correlations
\cite{NK14, IH07}.  
The question is where we may find any neutron-halo nuclei 
heavier than $^{37}$Mg.  The Coulomb and nuclear break-up data 
on the nucleus $^{37}$Mg were
interpreted as the one-neutron 
deformed p-wave halo \cite{NK14} with the neutron separation
energy (S$_{n}$) of 0.22$^{+0.12}_{-0.09}$ MeV.   
In the case of neutron p-wave halo the
S$_{n}$-values should be much smaller than, say, 1 MeV in order to clearly 
detect 
the characteristic feature of halo.  In contrast, in the case of neutron 
s-wave halo the presence of neutron halo can be detected  
for S$_n$-values appreciably larger than 1 MeV.  
Since S$_{n}$-values of possible halo nuclei not yet found 
in the region of neutron drip-line 
cannot be guessed so precisely and, thus, a successful prediction of  
neutron (spherical or deformed) 
p-wave halo must have very good luck due to the narrow range of 
S$_{n}$-values for the realization of halo phenomena,   
in the following I try mainly to look for
possible spherical or deformed s-wave halo nuclei heavier than $^{37}$Mg.    
In the present work only axially-symmetric quadrupole deformation 
is considered as deformation.
 
If I limit myself to spherical nuclei, the $\ell = 0$ halo can occur 
only around the nuclei with very particular neutron numbers,  
of which the neutron Fermi level is placed 
around neutron s$_{1/2}$ orbits.   One-particle s$_{1/2}$ orbit appears only
once in every N$_{ho}$ = even major shell, where N$_{ho}$ expresses 
the principal
quantum-number of the harmonic oscillator (ho) potential. 
The appearance may occur in nuclei around the neutron number N=51 
where the $3s_{1/2}$ orbit should be
filled (see Fig.3 of Ref. \cite{IH12}) by the 51st neutron in nuclei 
close to neutron-drip-line.     
In contrast, if nuclei are deformed, expressing the angular-momentum component
of the particle along the symmetry axis by $\Omega$ all
$\Omega^{\pi}$ = 1/2$^+$ levels have, in principle, a chance to become 
$\ell$=0 halo \cite{IH03}. This is because all 
$\Omega^{\pi}$ = 1/2$^{+}$ orbits acquire
$\ell$=0 components induced by the deformation 
and especially because the probability of the $\ell$=0
component in the bound one-particle 
wave-functions of neutron $\Omega^{\pi}$ = 1/2$^+$ orbits
increases strongly as the one-particle energy 
$\varepsilon_{\Omega^{\pi}=1/2^{+}}$ approaches zero \cite{TM97, IH04}.  

It is noted that the possibility of some isotopes in a certain region of the
neutron number to be deformed depends strongly on the proton number.  For
example, when the evolution of 2$_{1}$$^{+}$ energies in Ti, Cr and Fe isotopes
for the neutron-number varying from 34 towards 50 is examined, 
it is seen that 
$_{24}$Cr or $_{26}$Fe isotopes 
are easy to be deformed while $_{22}$Ti isotopes do not seem
to be so.  For example, see Fig. 9 of Ref. \cite{AG16}.  

Since for a given proton number Z 
the neutron separation energy of odd-N nuclei approaches zero
before even-N nuclei reach the neutron drip-line mainly due   
to the pairing-blocking effect in odd-N nuclei,    
it will be
experimentally easier to find neutron-halo nuclei among odd-N isotopes.
For example, while $^{37}$Mg$_{25}$ is found to be 
a deformed p-wave halo nucleus and
$^{39}$Mg$_{27}$ lies outside the neutron drip-line, both $^{38}$Mg$_{26}$ and
$^{40}$Mg$_{28}$ lie inside the neutron drip-line.
Moreover, generally speaking, the scheme of low-lying levels 
in odd-N nuclei would
directly show the underlying neutron shell-structure.   
Thus, in the present work I limit myself to the phenomena of one-neutron halo 
in nuclei with even-Z and odd-N   
in the region heavier than $^{37}$Mg, keeping the nuclei 
as light as possible.  
Partly because no quantitative information 
on the size of pair correlation in
nuclei around the relevant 
neutron drip-line is available, in the following I present the analysis of 
the wave function of the last-odd neutron in odd-N nuclei 
which is obtained by 
neglecting the pair correlation.   
Even in the presence of neutron pair-correlation 
in neighboring even-N nuclei 
the wave function of the last-odd neutron in the ground state of
odd-N nuclei can be studied in this way, at least in the first approximation.    

In Sec. II the model used and the result of search for even-Z odd-N nuclei 
with possible  
neutron halo are presented, while conclusions and discussions are given in Sec.
III.

\section{Model and candidates for even-Z odd-N nuclei with possible  
neutron halo}  
The basic points of the model are very similar to those used in Refs. 
\cite{IH12, IH14}. I use Woods-Saxon potentials with the standard parameters,
which are described on p. 239 of Ref. \cite{BM69}, unless   
I mention specifically some parameters.  
While both weakly-bound and resonant one-particle energies of neutrons in
neutron-rich nuclei are 
examined in the present work, the major part of the nuclear potentials for those
neutrons is produced by protons, which are deeply bound in the case of
neutron-rich nuclei.   Therefore, it may not be a surprise that the parameters
of the Woods-Saxon potential which were adjusted to various data on stable
nuclei in Ref. \cite{BM69} could satisfactorily work in the analysis of the
data on neutrons in neutron-rich nuclei \cite{MT14, TN09, NK14}.   

The way of calculating both bound and resonant one-particle energies 
in a spherical potential is straightforward 
and can be found in various textbooks.  
In a deformed potential the calculation of energies of bound neutrons is an
eigenvalue problem by solving 
in coordinate system the coupled differential equations obtained from
the Schr\"{o}dinger equation, together with the asymptotic
behavior of bound wave-functions in respective ($\ell, j$) channels 
for r $\rightarrow
\infty$.   On the other hand, one-particle resonant energies are calculated by
solving the coupled differential equations in coordinate space using the
asymptotic behavior of scattering states in respective ($\ell, j$) 
channels and are defined
as the energies, at which one of the eigenphases increases through $\pi$/2 as
the one-particle energy increases \cite{IH05}.   
One-particle resonance is absent if none of
the eigenphases increase through $\pi$/2 as the energy increases.  
In the limit of $\beta \rightarrow 0$ the one-particle resonances defined for
deformed potentials in terms of eigenphase, of course, agree with 
the one-particle resonances in spherical potentials defined 
in terms of phase shift.

The mean deformation of relevant nuclei is not microscopically calculated in the
present article, because a reliable effective interaction to be used in
Hartree-Fock (HF) calculations around neutron-drip-line is not yet available. 
Instead, I use the knowledge that a large energy gap in one-particle spectra for
some deformation as a function of deformation (Nilsson
diagram) indicates the stability of the system for the deformation and the
particle number.  Furthermore, I present the result only for prolate shape,
because the number of medium-heavy even-even nuclei, of which the ground state
is observed to be oblate is so few, though the overwhelming prolate dominance is
so far not really understood.   It is also noted that there is no trivial reason
why neutron-drip-line nuclei would prefer oblate shape, compared with nuclei
around the stability line.    

In the present work I must explore the probabilities of the $\ell$ = 0 or
$\ell$ = 1 component in the wave function of the last-odd neutron in the
laboratory system.  
When the even-even core configuration in a deformed odd-N nucleus 
is restricted to the ground state of the
neighboring deformed even-even nucleus which has I = K = 0, 
the probabilities of ($\ell, j$) components of the wave
function of the last-odd neutron in the deformed odd-N nucleus are 
obtained, in a good approximation,
from the angular-momentum projection of one-particle wave-functions of the
odd-neutron in corresponding deformed potentials.   The good approximation is
well known from the analysis of one-nucleon transfer reactions such as (d,p) and
(p,d) on the deformed ground-state of even-even nuclei, when proper kinematic
factors unique in respective reactions are taken into account.      
See, for example, Chapter 5-3 of Ref. \cite{BM75}.  
In reality, the even-even core configurations other than the 
ground-state configuration  may contribute to the ground-state 
wave-functions of odd-N
nuclei.   However, the contributions by those other configurations, for example
the lowest-lying (K=0, I=2) configuration or other excited core-configurations
mixed by rotational perturbation are pretty small in the ground and band-head
state of the neighboring odd-N nuclei.  Therefore, in the following when I look
for possible deformed halo nuclei, 
I am satisfied by finding one-particle levels of the
last-odd neutron, of which the wave functions in deformed potentials contain an
appreciable amount of $\ell = 0$ or $\ell = 1$ component.  In contrast, 
in spherical vibrational 
nuclei this approximation of the ground state of odd-A nuclei consisting
of the one-particle picture of the last-odd particle and 
the ground state of the neighboring even-even nuclei 
can be a much poorer approximation.

\subsection{even-Z odd-N nuclei around N=50 }
First, I figure out the level scheme of neutron one-particle levels as a
function of axially-symmetric quadrupole deformation for the nucleus, which can
be one of good candidates for deformed s-wave halo in the region of the neutron
number 50.  
Then, I try to find the possible 
lightest $\ell$ = 0 neutron-halo nuclei heavier than $^{37}$Mg, 
using the data on S$_{n}$-values of even-Z odd-N nuclei listed in Ref. 
\cite{BNL}.  
Some S$_{n}$ values in Ref. \cite{BNL} are ''evaluated'' values, which I also
use though those evaluated S$_{n}$-values have certainly an ambiguity that is
difficult to be estimated.    
Using those evaluated S$_{n}$-values is the best which I can do
for the moment.   In the following I mark those evaluated S$_{n}$-values 
by *.   
   
First of all, it is clear that when I limit myself to spherical shape  
the possible $\ell$ = 0 neutron halo which I look for 
must have a considerable amount of 
the neutron 3s$_{1/2}$ one-particle component.   In
the shell-structure of spherical stable nuclei the 3s$_{1/2}$ level is known to
appear in the second half of the major shell consisting of the neutron number 
N = 51-82.   In contrast, in the
case of very weakly-bound neutrons the 3s$_{1/2}$ level can be the lowest
one-particle level among the j-levels in the major shell consisting of 
N = 51-82 \cite{IH12}.    
In Fig.1 the neutron one-particle levels calculated in the Woods-Saxon
potential, namely both eigenenergies of bound neutrons 
for $\varepsilon_{\Omega} < 0$ and one-particle resonant energies for 
$\varepsilon_{\Omega} > 0$, are shown as a function of axially-symmetric 
quadrupole-deformation parameter $\beta$.  The 
parameters of the potential are taken from those  
for the nucleus $^{76}_{26}$Fe$_{50}$ except for the depth of the Woods-Saxon
potential, which is reduced by 1.6 MeV, and the strength of the spin-orbit
potential, which is about 80 percent of the standard value \cite{BM69}.  
The reduction of the depth by 1.6 MeV is made in order to make the 3s$_{1/2}$
level at $\beta$=0 being bound, while 
the 80 percent spin-orbit 
strength is chosen in order just to avoid the complicated
level-crossing around $\beta \approx 0$.   The resulting shell-structure for
$\beta > 0$ to be discussed in the following is not really changed by these
changes of the potential parameters.   
The lowest-lying $\Omega^{\pi}$ = 1/2$^{+}$ level in Fig. 1 coming from the
spherical N$_{ho}$ = 4 shell, which is denoted by [440 1/2] and 
is connected to the 1g$_{9/2}$ level at
$\beta$ = 0, does not contain an appreciable amount of $\ell = 0$ component for
realistic values of quadrupole deformation $\beta < 0.5$, simply because of
the vanishing matrix-element of the $Y_{20}$ operator between the 
1g$_{9/2}$ and 3s$_{1/2}$ orbits.  The quadrupole coupling between the two
orbits becomes effective mainly via the neighboring 2d$_{5/2}$ orbit, which
couples with both the 1g$_{9/2}$ and 3s$_{1/2}$ orbits by the $Y_{20}$ operator.     
For example, the probability of $\ell$=0 component in the wave function of the
lowest-lying $\Omega^{\pi}$ = 1/2$^{+}$ level in Fig. 1 is 0.014, 0.035 and     
0.041 for $\beta = 0.2$, 0.4 and 0.6, respectively.

Thus, except for the case that one-particle energy is extremely small, 
among the levels
coming from the N$_{ho}$ = 4 major shell in Fig. 1 
the lowest-lying  $\Omega^{\pi}$ = 1/2$^{+}$ neutron level, 
which for realistic
values of $\beta (< 0.5)$ contains an
appreciable amount of the $\ell = 0$ component, is the second (or
third) lowest $\Omega^{\pi}$ = 1/2$^{+}$ level    
indicated as the [431 1/2] (or [420 1/2]) level.   
Limiting myself to spherical or prolately deformed shape, 
the neutron numbers of
which the Fermi level may be placed on the second or third lowest
$\Omega^{\pi}$ = 1/2$^{+}$ level in Fig. 1 are N=47 with the [431 1/2] level 
for $0.18 < \beta < 0.38$,  
N=49 with the [431 1/2] level for $0.10 < \beta < 0.18$ and with the 
[420 1/2] level for $0.32 < \beta < 0.52$,   
and N=51 with the [431 1/2] level for $0 \leq \beta < 0.10$ and with the 
[420 1/2] level for $0.19 < \beta < 0.32$.  
In all the bands constructed based on those $\Omega^{\pi}$ = 1/2$^{+}$ 
one-particle levels the decoupling parameters $a$ are found to be 
in the range of 
$-1 < a < +4$ for the relevant deformations and, thus, the band-head states in
the laboratory system have
the spin-parity 1/2$^{+}$.   

For Ni-isotopes the S$_{n}$-values of $^{71}_{28}$Ni$_{43}$, 
$^{73}_{28}$Ni$_{45}$, $^{75}_{28}$Ni$_{47}$ and
$^{77}_{28}$Ni$_{49}$ are 4.3, 4.0, 3.9$^{*}$ and 3.2$^{*}$ MeV, respectively.  
Thus, though
$^{79}_{28}$Ni$_{51}$ would lie inside the neutron drip-line and be presumably
spherical, the S$_n$-value would not be small enough to expect that the nucleus
may show a possible halo.  
For Ti-isotopes the S$_{n}$-values of $^{55}_{22}$Ti$_{33}$, 
$^{57}_{22}$Ti$_{35}$,    
$^{59}_{22}$Ti$_{37}$,  $^{61}_{22}$Ti$_{39}$  and
$^{63}_{22}$Ti$_{41}$ are 4.1, 2.7, 2.6$^{*}$, 2.1$^{*}$ 
and 1.3$^{*}$ MeV, respectively.  
Therefore, the odd-N
nucleus $^{69}_{22}$Ti$_{47}$ may be 
expected to lie outside the neutron drip-line.   
Then, examining available data on nuclei towards the possible neutron
drip-line, promising candidates for s-wave neutron-halo nuclei are 
$^{71}_{24}$Cr$_{47}$, $^{73}_{24}$Cr$_{49}$, $^{75}_{24}$Cr$_{51}$ and
$^{77}_{26}$Fe$_{51}$, in the case that those nuclei lie inside the neutron 
drip-line.   
(The two neutron separation energies S$_{2n}$ of those nuclei are 
expected to be positive so that the nuclei are stable against 2n
emission.)      
The reason is the following: The heaviest member of
Cr-isotopes so far found was $^{70}_{24}$Cr$_{46}$ \cite{OBT13} while 
S$_n$-values of $^{61}_{24}$Cr$_{37}$,   
$^{63}_{24}$Cr$_{39}$,  $^{65}_{24}$Cr$_{41}$ and $^{67}_{24}$Cr$_{43}$ are
4.0, 2.9, 2.6$^{*}$ and 2.0$^{*}$ MeV, respectively.   
For Fe-isotopes the presence of $^{75}_{26}$Fe$_{49}$ was
observed at MSU \cite{OBT13}, while S$_{n}$-values of $^{65}_{26}$Fe$_{39}$, 
$^{67}_{26}$Fe$_{41}$,  $^{69}_{26}$Fe$_{43}$ and
$^{71}_{26}$Fe$_{45}$ are 4.3, 4.0, 3.3$^{*}$ and 2.8$^{*}$ MeV, respectively.
In short, the S$_{n}$-values of $^{71}$Cr, $^{73}$Cr, $^{75}$Cr 
and $^{77}$Fe may be
less than 1 MeV, in the case that they are not negative.     

It is noted that the
shell-structure for $\beta > 0$ consisting of both weakly-bound
levels and low-lying one-particle resonances in Fig. 1 remains 
almost quantitatively unchanged when the depth of the Woods-Saxon potential is
changed by, say less than 1.5 MeV so that some of one-particle levels 
are changed 
from bound states to one-particle resonances or vice versa.  Therefore, 
Fig. 1 which is
for the system [$^{76}$Fe + n] is used to explore the possibility of deformed
s-wave halo in nuclei with N = 47, 49 and 51 which have weakly-bound neutrons.  

The similarity of the neutron shell-structure above N=40 to that above N=20 
for respective 
neutron-rich nuclei is sometimes pointed out, for example, see Ref. \cite{AG16}.
The latter leads to the region of deformed nuclei called ''island of
inversion''.   For nuclei with weakly-bound neutrons the spherical 
energy gap at N=50
becomes smaller as seen in Fig. 1    
and, moreover, four one-particle levels, 1g$_{9/2}$, 3s$_{1/2}$, 2d$_{5/2}$ and 
2d$_{3/2}$, occur within a few MeV at $\beta$ = 0.   
Consequently, the ground 
states of those nuclei with N $\approx$ 41-51 may be deformed, if the 
proton numbers of respective nuclei allow the deformation.
The neutron number 51 is the sum of N=40 and a half of the neutron number 22,
which can be accommodated in the close-lying shells consisting of 
1g$_{9/2}$, 3s$_{1/2}$, 2d$_{5/2}$ and 2d$_{3/2}$.    
In this respect, it 
is interesting to note that in the recent shell-model calculations 
in \cite{FN16} the
nuclei $^{74}_{24}$Cr$_{50}$ and $^{76}_{26}$Fe$_{50}$ are
predicted to be deformed, though in the calculation of Ref. \cite{FN16}
harmonic-oscillator wave-functions and ESPEs (Effective Single Particle
Energies), which are adjusted based on the knowledge of single-particle energies
in nuclei far away from neutron-drip-line, 
are used. That means, the important effect of very
weak bindings of neutrons on the neutron single-particle energies 
is not taken into account in Ref. \cite{FN16}.

\subsection{even-Z odd-N nuclei around N=34}
In this subsection I describe a possibility of halo phenomena related to the
nucleus $^{53}_{18}$Ar$_{35}$.  The presence of $^{53}$Ar inside the neutron
drip-line was reported in MSU 
\cite{OBT09}.  The proton number Z = 18 does not seem to particularly favor  
prolate deformation, though some Ar isotopes such as $^{48}_{18}$Ar$_{30}$ can
be interpreted as being deformed.  

The heaviest member of Ar-isotopes so far found was $^{53}$Ar, while 
S$_n$-values of $^{45}_{18}$Ar$_{27}$,  $^{47}_{18}$Ar$_{29}$, 
$^{49}_{18}$Ar$_{31}$ and
$^{51}_{18}$Ar$_{33}$ are 5.17(0.002), 3.55(0.08), 3.60(1.41) from Ref.
\cite{ZM15} and 1.0$^{*}$ MeV, respectively.   
Thus, the
S$_n$-value of $^{53}$Ar is expected to be small.  If the nucleus is
spherical or slightly oblate, the spin-parity of 
the ground state is likely to be 5/2$^{-}$.   
However, in the following it is shown 
that if the nucleus is prolately deformed and the S$_n$-value is smaller than, 
say several hundreds keV, there may be a chance for the ground state 
to show deformed s-wave or p-wave halo phenomena.  

As seen in Fig. 1, the possible lowest-lying $\Omega^{\pi}$ = 1/2$^{+}$ 
one-particle level 
on the prolate side coming from the N$_{ho}$ = 4 
major shell is the one connected to the
1g$_{9/2}$ level at $\beta$ = 0.  As described in the previous
subsection, this lowest-lying
$\Omega^{\pi}$ = 1/2$^{+}$ level contains usually only less than 
a few percent probability of
$\ell$=0 component in the range of realistic quadrupole deformation, 
say $\beta <$ 0.5.   The only exception is the case that this 
$\Omega^{\pi}$ = 1/2$^{+}$ level has a very small binding energy, 
say $|\varepsilon_{\Omega}|$ smaller than a few hundred keV.  
As described in \cite{TM97, IH04} the
probability of $\ell$ = 0 component in the one-particle neutron wave-function
with $\Omega^{\pi}$ = 1/2$^{+}$
approaches unity in the limit of $|\varepsilon_{\Omega}|$ $\rightarrow 0$.  
The value of $|\varepsilon_{\Omega}|$, at which the probability of $\ell$ = 0  
component starts to strongly increase, 
depends on respective $\Omega^{\pi}$ = 1/2$^{+}$ orbits.

In Fig. 2 neutron one-particle energies are plotted as a function of
axially-symmetric quadrupole deformation parameter $\beta$, taking the
standard parameters of Ref. \cite{BM69} for the system of 
[$^{52}_{18}$Ar$_{34}$ + n].  
Each orbit is doubly degenerate.  Thus, if I take literally this figure, the
35th neutron in $^{53}$Ar occupies the orbit with $\Omega^{\pi}$ = 5/2$^{-}$ for
$-0.16 < \beta < 0$, $\Omega^{\pi}$ = 1/2$^{-}$ for $0 < \beta < 0.13$, 
$\Omega^{\pi}$ = 3/2$^{-}$ for $0.13 < \beta < 0.27$ and 
$\Omega^{\pi}$ = 1/2$^{+}$ for $0.27 < \beta < 0.41$.  
When I examine the wave functions of neutrons, I find that for $\beta < 0.05$ 
the
main component of the $\Omega^{\pi}$ = 1/2$^{-}$ orbit will be
f$_{5/2}$, while that of the $\Omega^{\pi}$ = 3/2$^{-}$ orbit will be
f$_{5/2}$ for $\beta < 0.18$.  
Therefore, deformed p-wave halo may not be obtained for respective
$\beta$-values.  On the other hand, if the
S$_{n}$-value is larger than about 400 keV 
the probability of $\ell$=0 component in
the wave function of the lowest-lying 
$\Omega^{\pi}$ = 1/2$^{+}$ orbit is less than 0.10 and the
main components of the one-neutron wave-function become g$_{9/2}$ 
and d$_{5/2}$.  
Both the calculated decoupling parameter of the $\Omega^{\pi}$ = 1/2$^{-}$
level for $0 < \beta < 0.3$ and that of the $\Omega^{\pi}$ = 1/2$^{+}$ level
for $0.24 < \beta < 0.5$ are in the range of $-1 < a < +4$.   Thus, in the
absence of band-mixing the 
spin-parity of the band-head states based on those one-particle orbits with  
$\Omega$ = 1/2 
would be I$^{\pi}$ = 1/2$^{-}$ and I$^{\pi}$ =
1/2$^{+}$ in the range of respective deformations.   On the other hand, 
for $0 < \beta < 0.2$ the one-particle energies $|\varepsilon_{\Omega}|$ of the
very-weakly bound levels with $\Omega^{\pi}$ = 1/2$^{-}$ and 3/2$^{-}$ are 
so close that the two
configurations constructed based on the two respective one-particle levels 
may be strongly mixed 
in observed states.
It is rather difficult to say something more in the present
stage since no spectroscopic information on the nucleus $^{53}$Ar is available.
Therefore, I would state that the condition literally obtained from Fig. 2 is: 
the nucleus $^{53}$Ar is expected to show deformed
p-wave halo 
if it has a moderate deformation of $\beta \approx 0.20$-0.25 while it will
show deformed
s-wave halo for $\beta \approx$ 0.25-0.40 if it has an S$_{n}$-value 
smaller than a few hundred keV.      

I add just one comment on the decoupling parameters of 
very weakly-bound $\Omega$ = 1/2 one-particle levels. 
In the lowest-lying $\Omega^{\pi}$ = 1/2$^{+}$ 
level in Fig. 2 connected to the
1g$_{9/2}$ level at $\beta$=0 the decoupling parameter generally increases as 
$\beta (>0) \rightarrow 0$ and approaches +5 which is the value of 
the decoupling parameter for a j=9/2 particle.   However, 
as $\beta (>0.24) \rightarrow 0.24$ in Fig. 2, 
the value of $|\varepsilon_{\Omega} (<0)|$ approaches zero 
and the probability of 
the $\ell = 0$ component in
the one-particle wave-function steeply increases and 
approaches unity.  Then, the decoupling parameter
of course approaches the value of +1, which is the value for a j=1/2 particle.   

Though in order for the ground state of $^{53}$Ar to show halo phenomena
the conditions described above must be luckily satisfied, 
it would be very interesting if one can find the halo phenomena 
in the nucleus because it may exhibit exciting  
quantum-mechanical features such as deformation, deformed shell-structure of
weakly-bound neutrons, the 
possible strong increase of $\ell$=0 component in the wave-functions of 
very weakly-bound $\Omega^{\pi}$ = 1/2$^{+}$ neutrons, and so on.      
Since the presence and a presumably small S$_{n}$-value of the nucleus 
$^{53}$Ar are already guaranteed,    
I hope a proper experiment will be 
carried out in the near future.

\section{Conclusions and discussions}
Recognizing that the nucleus $^{37}$Mg is so far the heaviest halo nucleus which
was experimentally found, one may wonder whether the halo phenomena can be
found also in very heavy nuclei.  Then, a question is which nucleus heavier
than $^{37}$Mg will be next found to 
show the halo property.   Using the shell structure unique
in nuclei with weakly-bound neutrons, I have tried to look for 
possible neutron-halo nuclei heavier than $^{37}$Mg to be next studied.   
Candidates for such deformed s-wave halo nuclei are those with the neutron
number N=47, 49 and 51.  
Examining available S$_{n}$-values of even-Z odd-N isotopes, promising nuclei
are $^{71}_{24}$Cr$_{47}$, $^{73}_{24}$Cr$_{49}$, $^{75}_{24}$Cr$_{51}$ and
$^{77}_{26}$Fe$_{51}$.   
The $\ell$=0 components in the wave functions of those
possible halo neutrons come mainly 
from the 3s$_{1/2}$ level.  And, one-neutron level
which contains an appreciable amount of the $\ell = 0$ component 
is, except for the 
very weakly binding case of the lowest-lying $\Omega^{\pi}$ = 1/2$^{+}$ level,  
the second (or third)-lowest $\Omega^{\pi}$ = 1/2$^{+}$ 
neutron level among the levels coming from 
the spherical N$_{ho}$=4 major shell.   
Those promising candidates for halo nuclei are expected to be deformed 
because in
the spherical limit the N=50 energy gap is narrowed and a set of close-lying
one-particle levels, 1g$_{9/2}$, 3s$_{1/2}$, 2d$_{5/2}$ and 2d$_{3/2}$, appear. 

In other words, the possible observation of an s-wave halo in the above N = 47,
49 and/or 51 nuclei directly suggests the shell-structure unique in very 
weakly-bound neutrons, such as the disappearance of N = 50 magic number 
and the very lowering of the 3s$_{1/2}$ level inside the major shell 
50 $<$ N $<$ 82.   
   
Proper mean-field calculations such as HF calculations to obtain 
nuclear shape have not properly been 
carried out for those neutron-rich nuclei with weakly-bound neutrons mainly
because of the lack of the proper HF  
two-body interaction which has been 
established for those nuclei.  
Under these circumstances I have used the fact that the presence of a few
close-lying one-particle levels in the spherical limit is known to lead to the
prolately deformed shape of nuclei when neutrons start to fill 
the first half of the close-lying levels (Jahn-Teller effect), though the
possible presence of pair-correlation which acts against the tendency of 
deformation may delay the occurrence of deformation in the beginning of the
first half.   Examples are the
deformation of stable rare-earth nuclei with 90$\leq$N$\leq$112 \cite{BM75}  
and the deformation of neutron-rich Mg isotopes \cite{IH07}.  
By the way, to my knowledge 
the reason why very few oblately deformed nuclei with neutrons
filling in the second half of a major shell (or close-lying shells)   
are realized has not yet really been pinned down.   

In the present work I had to use the ''evaluated'' (or extrapolated) 
S$_{n}$-values in Ref. \cite{BNL} for some neutron-rich nuclei, which are often 
crucial to predict possible halo nuclei.  There is no other more-reliable way,
therefore, I could not avoid that  
the resulting conclusion
contains the corresponding ambiguity.   

The basis to expect that the nucleus $^{53}$Ar is 
prolately deformed is certainly weaker,
however, the interpretation of the observed prolate deformation of nuclei 
such as $^{31}_{12}$Mg$_{19}$  \cite{GN05} 
or $^{29}_{10}$Ne$_{19}$ \cite{NK16} in the region of island of inversion 
encourages the expectation.   
Namely, as $\beta (>0) \rightarrow$ large, one-particle levels with 
$\Omega^{\pi}$ = 1/2$^{+}$ and $\Omega^{\pi}$ = 3/2$^{+}$ originating from the
high-j (1g$_{9/2}$ in the present case) 
shell energetically come down steeply in Fig. 2, 
and the filling of those one-particle levels may help 
for the relevant nuclei to acquire prolate deformation.   

To my knowledge, in heavier stable nuclei the pair-correlation becomes important
almost whenever deformation is likely to occur.   
In the present work one-neutron wave-functions in the absence of
pair-correlation are used since the neutron-halo phenomena in the ground state
of odd-N nuclei are 
explored.   It may be interesting to find how halo phenomena in heavier
odd-N nuclei are affected by a possible strong pair-correlation in the
even-even core nuclei.   

The author expresses her thanks to professor T. Nakamura for his comments.

\newpage

\noindent
{\bf\large Figure captions}\\
\begin{description}
\item[{\rm Figure 1 :}]
Calculated one-particle energies for neutrons in the potential given by the
core nucleus $^{76}_{26}$Fe$_{50}$ as a function of axially-symmetric 
quadrupole deformation parameter $\beta$.
Bound one-particle energies at $\beta$ = 0 are $-$6.56, $-$6.14, $-$5.39,
$-$1.53, and $-$0.05 MeV
for the $2p_{3/2}$, $1f_{5/2}$, $2p_{1/2}$, $1g_{9/2}$ and $3s_{1/2}$ levels 
respectively, while
one-particle resonant $2d_{5/2}$, $2d_{3/2}$ and $1g_{7/2}$ levels are obtained 
at 0.25, 1.62 and 3.12 MeV (denoted by filled circles) 
with the width of 0.024, 1,92 and
0.13 MeV, respectively.  
The $\Omega^{\pi}$ = 1/2$^{+}$ resonant level connected to the 2d$_{5/2}$ level
at $\beta$ = 0 cannot be obtained as a one-particle resonant level for 
$\beta > 0.007$.  
Three bound $\Omega^{\pi}$ = 1/2$^{+}$ levels are denoted by the asymptotic
quantum-numbers on the prolate side \cite{BM75}, [$N n_z \Lambda \Omega$] = 
[440 1/2], [431 1/2] and [420 1/2], although the wave-functions
of those levels plotted in this figure are quite different from those
expressed by the asymptotic quantum-numbers.   
The neutron numbers, 40, 48 and 50, which
are obtained by filling all lower-lying levels, are indicated with circles, for
reference.     
For simplicity, calculated widths of one-particle resonant levels are not shown.
The parity of levels can be seen from
the $\ell$-values denoted at $\beta$ = 0; $\pi$ = $(-1)^{\ell}$.
One-particle resonant energies for $\beta \neq 0$ are not always 
plotted if they are
not important for the present discussions. 
\end{description}

\begin{description}
\item[{\rm Figure 2 :}]
Calculated one-particle energies for neutrons in the potential given by 
the core nucleus $^{52}_{18}$Ar$_{34}$ 
as a function of axially-symmetric quadrupole deformation.
Bound one-particle energies at $\beta$ = 0 are $-$5.99, $-$3.18, $-$1.63 and 
$-$0.31 MeV for the $1f_{7/2}$, $2p_{3/2}$, $2p_{1/2}$  
and $1f_{5/2}$ levels, respectively, while one-particle resonant $1g_{9/2}$
level is obtained at 2.69 MeV with the width of 0.061 MeV.  
The $2d_{5/2}$ level is not obtained as a one-particle
resonant level for this potential. 
The neutron numbers, 28 and 34, which are obtained 
by filling all lower-lying
levels, are indicated with circles. 
One-particle resonant energies for $\beta \neq 0$ are not always 
plotted if they are
not important for the present discussions. 

\end{description}


\vspace{2cm}
\begin{thebibliography}{99}
\bibitem{IT85} I. Tanihata {\it et al.}, Phys. Rev. Lett. {\bf 55}, 2676
(1985).
\bibitem{MT14} For example, M. Takechi {\it et al.}, Phys. Rev. C {\bf 90}, 
061305 (2014).
\bibitem{TN09} T. Nakamura {\it et al.}, Phys. Rev. Lett. {\bf 103}, 262501 
(2009); Phys. Rev. Lett. {\bf 112}, 142501 (2014).
\bibitem{NK14} N. Kobayashi {\it et al.}, Phys. Rev. Lett. {\bf 112}, 242501 
(2014).  
\bibitem{IH07} I. Hamamoto, Phys. Rev. C {\bf 76}, 054319 (2007).
\bibitem{IH12} I. Hamamoto, Phys. Rev. C {\bf 85}, 064329 (2012).
\bibitem{IH03} I. Hamamoto and B. R. Mottelson, C. R. Phys. {\bf 4}, 433 (2003).
\bibitem{TM97} T. Misu, W. Nazarewicz and S.\AA berg, Nucl. Phys. {\bf A614}, 
44 (1997).   
\bibitem{IH04} I. Hamamoto, Phys. Rev. C {\bf 69}, 041306(R) (2004).   
\bibitem{AG16} A. Gade and B. Sherrill, Phys. Scri. {\bf 91}, 053003 (2016).
\bibitem{IH14} I. Hamamoto, Phys. Rev. C {\bf 89}, 057301 (2014).
\bibitem{BM69} A. Bohr and B. R. Mottelson, {\it Nuclear Structure\/} (Benjamin,
Reading, MA, 1969), Vol. I, p.239.
\bibitem{IH05} I. Hamamoto, Phys. Rev. C {\bf 72}, 024301 (2005); {\bf 73},
064308 (2006).
\bibitem{BM75} A. Bohr and B. R. Mottelson, {\it Nuclear Structure\/} (Benjamin,
Reading, MA, 1975), Vol.II.  
\bibitem{BNL} National Nuclear Data Center, http://www.nndc.bnl.gov.  
\bibitem{OBT13} O. B. Tarasov {\it et al.}, Phys. Rev. C {\bf 87}, 054612 
(2013).
\bibitem{FN16} F. Nowacki, A. Poves, E. Caurier and B. Bounthong,
arxiv:1605.05103v2 [nucl-th] 7Oct2016.   
\bibitem{OBT09} O. B. Tarasov {\it et al.}, Phys. Rev. Lett.  {\bf 102}, 142501 
(2009).
\bibitem{ZM15} Z. Meisel {\it et al.}, Phys. Rev. Lett.  {\bf 114}, 022501 
(2015).
\bibitem{GN05} G. Neyens {\it et al.}, Phys. Rev. Lett. {\bf 94}, 022501 (2005).
\bibitem{NK16} N. Kobayashi {\it et al.}, Phys. Rev. C {\bf 93}, 014613 (2016).  

\end{thebibliography}
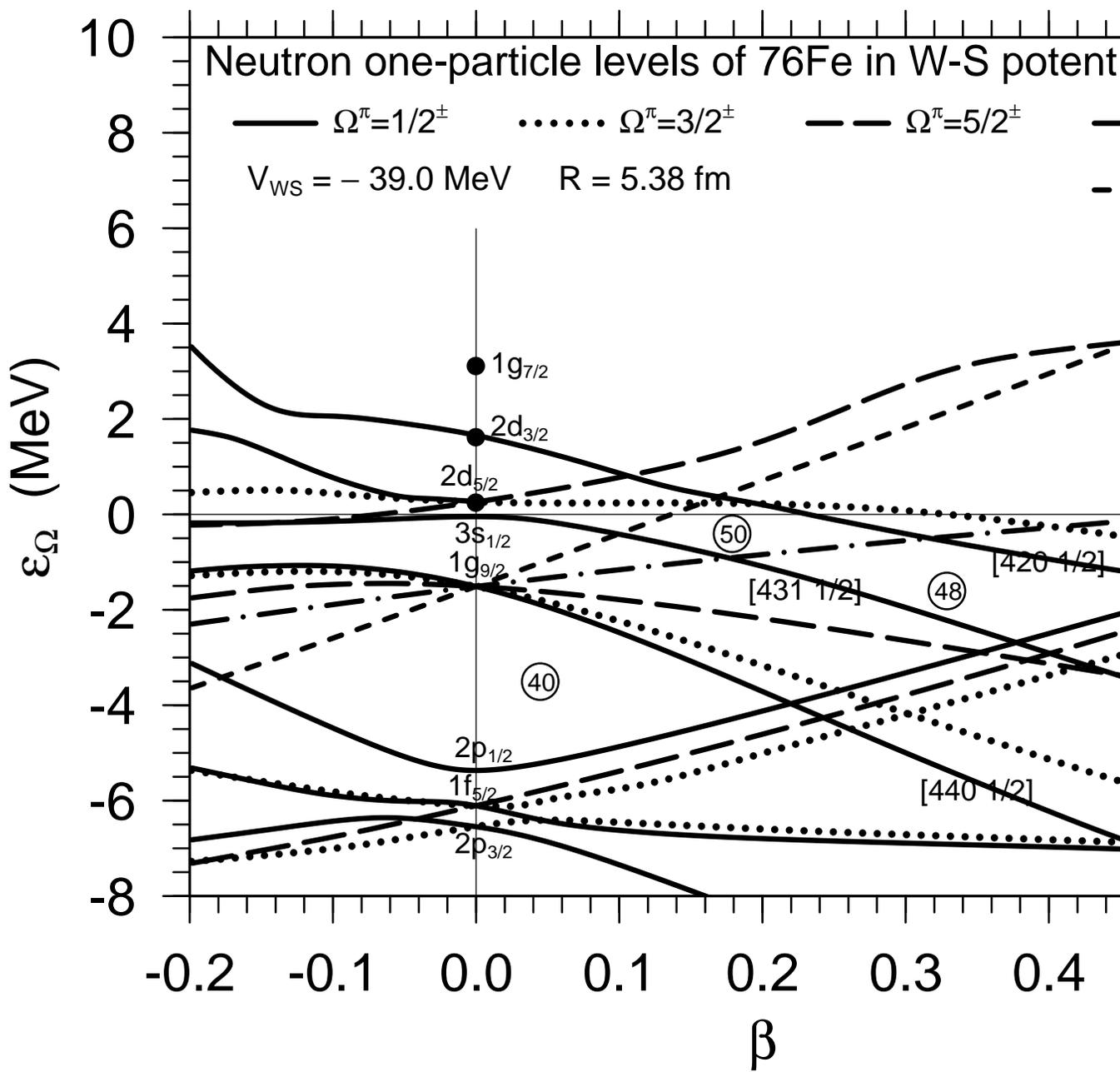
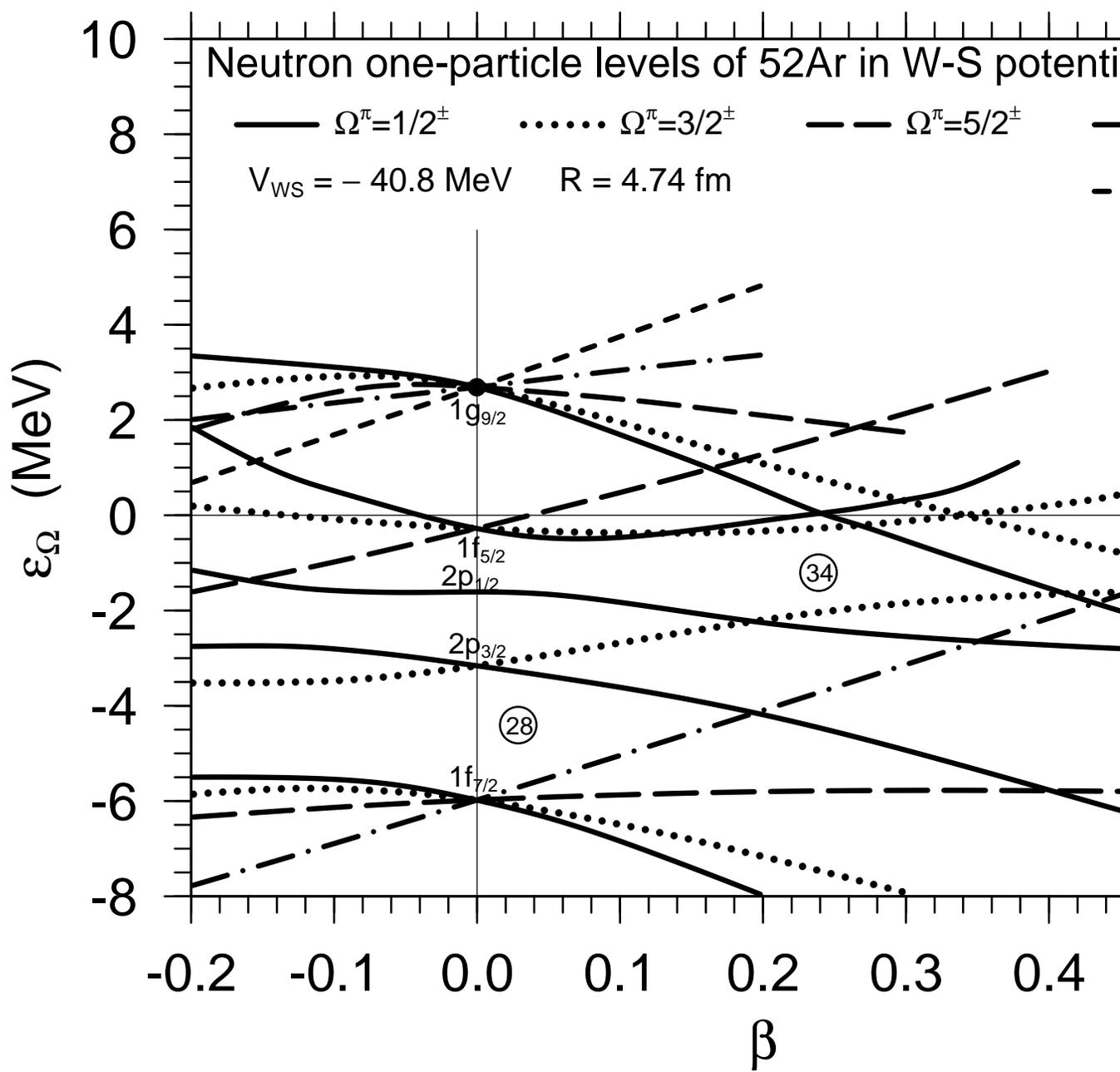
\end{document}